\documentclass[12pt]{article}

\begin{document}

\vspace*{1.0cm}

\begin{center}
{\large {\bf Relativistic Equation of State of Nuclear Matter \\
        for Supernova and Neutron Star\\ }  }
\vspace*{0.5cm}
H. Shen$^{a,b,}$\footnote{e-mail: shen@rcnp.osaka-u.ac.jp},
H. Toki$^{a,c,}$\footnote{e-mail: toki@rcnp.osaka-u.ac.jp},
K. Oyamatsu$^{c,d,}$\footnote{e-mail: oyak@postman.riken.go.jp}
and K. Sumiyoshi$^{c,}$\footnote{e-mail: sumi@postman.riken.go.jp} \\
\end{center}
      $^{a}$Reaserch Center for Nuclear Physics(RCNP), Osaka University,
           Ibaraki, Osaka 567, Japan \\
      $^{b}$Department of Physics, Nankai University, Tianjin 300071, China \\
      $^{c}$The Institute of Physical and Chemical Research (RIKEN),
            Wako, Saitama 351, Japan \\
      $^{d}$Department of Energy Engineering and Science, Nagoya University,
            Nagoya 464-01, Japan \\

\vspace*{0.5cm}
\begin{abstract}
  We construct the equation of state (EOS) of nuclear matter using the
relativistic mean field (RMF) theory in the wide density, temperature
range with various proton fractions for the use of supernova simulation
and the neutron star calculations.
We first construct the EOS of homogeneous nuclear matter.
We use then the Thomas-Fermi approximation to describe inhomogeneous
matter, where heavy nuclei are formed together with free nucleon gas.
We discuss the results on free energy, pressure and entropy in the
wide range of astrophysical interest.
As an example, we apply the resulting EOS on the neutron star properties
by using the Oppenheimer-Volkoff equation.
\end{abstract}



\newpage

\section{Introduction}

Supernova is a spectacular stellar phenomenon happening when massive stars
($M \geq 8M_{\odot}$) use up the nuclear fuel.
Due to the gravitational core collapse, supernova explosion may lead to
the formation of a neutron star or black hole in the central part of the
massive star.
In order to clarify the mechanism of the explosion and the whole
phenomena of supernova explosion, it is necessary to study
them by performing numerical simulations.
For this purpose, we need the equation of state (EOS) of nuclear matter in the
wide density and temperature range with various proton fractions,
which appear during the collapse and the explosion.

There have been several EOS worked out so far for supernova study
\cite{BV81,OS82,HW85,LLPR85,LS91}.
Most of them are constructed based on the non-relativistic Skyrme Hartree-Fock
(SHF) framework. Although some of these EOS are carefully constructed
in the form for the supernova simulations, there
are many places where the EOS are discontinuous or not available, and the
supernova
simulations stop there due to the lack of the completeness of the EOS
table.
Sometimes, they are not even documented and it is difficult to fix the
problem. Lattimer and Swesty worked out the EOS in the compressible
liquid drop model based on the non-relativistic framework \cite{LS91}.
This EOS table is almost the unique one,
with which many simulations are made at present.

Recently, there is a tremendous development in the description of nuclei
and nuclear matter in terms of the relativistic many body theory
\cite{BM90,LMB92,EO96}.
The relativistic Br\"uckner-Hartree-Fock (RBHF) theory provided the
saturation close to the experimental point in nuclear matter.
The relativistic effect provides the strong density dependent repulsion
over the non-relativistic results due to the three-body Z-graph process.
The RBHF theory provides also the strong spin-orbit splittings, which
are the fundamental quantities of the shell model.
Based on this finding of the RBHF theory, the phenomenological
model was developed using the relativistic mean field (RMF) theory
\cite{SW86,BT92,ST94,SST97}.
There are many works performed on nuclear ground states using the RMF theory.
The results compare very well with experimental data on not only stable
nuclei but also unstable nuclei, which are being obtained using radioactive
nuclear beams \cite{T95}.

Having success in describing nuclear properties using the RMF theory,
it is extremely important to work out the EOS for astrophysical purposes.
We have to work out consistent calculations for high density nuclear matter,
inhomogeneous nuclear matter and finite nuclei under the
wide range of density, proton fraction and temperature, which appear
inside neutron stars and supernovae.
For this purpose, we study the properties of dense matter
with both homogeneous and inhomogeneous distribution
in the RMF framework adopting the parameter set TM1 for the RMF lagrangian.
The RMF theory with the TM1 is known to provide
excellent properties of the ground states of heavy nuclei
including unstable nuclei \cite{ST94,HS97}.
The TM1 was also used  for the giant resonances
within the RPA formalism and demonstrated to provide good descriptions
of giant resonances \cite{MT97}.

For the description of inhomogeneous nuclear matter, which appears below
$\rho \sim \rho_0 /3 $ with $\rho_0$ being the normal nuclear matter density,
we use the Thomas-Fermi approximation \cite{O93}.
We assume that the inhomogeneous matter is composed of a lattice of spherical
nuclei. At finite temperatures, a proton gas, as well as a neutron gas, may
be present outside the nuclei.
We minimize the free energy of the system with respect to the cell volume,
proton and neutron radii and surface diffusenesses of the nucleus, and
densities of the outside neutron and proton gases.
We compare the calculated free energy with that of homogeneous matter
to determine the phase of dense matter.
This has to be done for each density, temperature and proton fraction
in the wide range necessary for the astrophysical use.

This paper is arranged as follows. In Sect.2, we describe the RMF theory and
the TM1 parameter set. In Sect.3, we introduce the Thomas-Fermi method to
be used with the Wigner-Seitz approximation. In Sect.4, we show the numerical
results. In Sect.5 we show the neutron star profiles using the EOS table
with the Oppenheimer-Volkoff equation. Sect.6 is devoted to the
conclusions and discussions.

\section{ Relativistic mean field theory}

We briefly decribe the relativistic mean field theory here.
All the details are to be referred to Ref.\cite{SW86,ST94,ST94AJ,SKT95}.
We start with the lagrangian given by
\begin{eqnarray}
{\cal L}_{RMF} & = & \bar{\psi}\left[i\gamma_{\mu}\partial^{\mu} -M
-g_{\sigma}\sigma-g_{\omega}\gamma_{\mu}\omega^{\mu}
-g_{\rho}\gamma_{\mu}\tau_a\rho^{a\mu}
-e\gamma_{\mu}\frac{1-\tau_3}{2}A^{\mu}\right]\psi  \\ \nonumber
 && +\frac{1}{2}\partial_{\mu}\sigma\partial^{\mu}\sigma
-\frac{1}{2}m^2_{\sigma}\sigma^2-\frac{1}{3}g_{2}\sigma^{3}
-\frac{1}{4}g_{3}\sigma^{4} \\ \nonumber
 && -\frac{1}{4}W_{\mu\nu}W^{\mu\nu}
+\frac{1}{2}m^2_{\omega}\omega_{\mu}\omega^{\mu}
+\frac{1}{4}c_{3}\left(\omega_{\mu}\omega^{\mu}\right)^2   \\ \nonumber
 && -\frac{1}{4}R^a_{\mu\nu}R^{a\mu\nu}
+\frac{1}{2}m^2_{\rho}\rho^a_{\mu}\rho^{a\mu}
-\frac{1}{4}F_{\mu\nu}F^{\mu\nu} ,
\end{eqnarray}
where
\begin{eqnarray}
W^{\mu\nu} & = & \partial^{\mu}\omega^{\nu}
               - \partial^{\nu}\omega^{\mu} ,     \\ \nonumber
R^{a\mu\nu} & = & \partial^{\mu}\rho^{a\nu}- \partial^{\nu}\rho^{a\mu}
 +g_{\rho}\epsilon^{abc}\rho^{b\mu}\rho^{c\nu} ,   \\ \nonumber
F^{\mu\nu} & = & \partial^{\mu}A^{\nu}
                -\partial^{\nu}A^{\mu}.
\end{eqnarray}
We adopt the lagrangian with the non-linear $\sigma$ and $\omega$ terms,
which was studied by Sugahara and Toki \cite{ST94}.
It was shown that the inclusion of the non-linear $\sigma$ terms is
essential to reproduce the properties of nuclei quantitatively and
a reasonable value for the incompressibility, while the
non-linear $\omega$ term was added in order to reproduce the density
dependence of the vector part of the self-energy of the nucleon obtained in
the RBHF theory.

The lagrangian contains the coupling constants, the meson masses, and the
self-coupling constants as parameters. We adopt the parameter set TM1,
which was determined in Ref.\cite{ST94} as the best one to reproduce
the properties of finite nuclei in the wide mass range in the periodic
table including neutron-rich nuclei.
They made the least-square fitting to the experimental data of the
binding energies and the charge radii of proton magic nuclei from Ca to Pb
including unstable Pb isotopes. The RMF theory with the TM1 parameter set
was also shown to reproduce satisfactory agreement with experimental data
in the studies of the nuclei with deformed configuration \cite{HS97} and the
giant
resonances within the RPA formalism \cite{MT97}.
With the TM1 parameter set, the symmetry energy is 36.9 MeV and
the incompressibility is 281 MeV.
The parameter set TM1 can be found in Ref.\cite{ST94,SKT95}.

The derivation of the Euler-Lagrange equations and the following
treatment for the EOS of homogeneous nuclear matter at finite temperature
is the same as the procedure described in Ref.\cite{ST94AJ,SKT95}.
The expressions for physical quantities such as energy density and
entropy, which are used in the Thomas-Fermi calculations, can be
found in \cite{ST94AJ}.
Properties of homogenous nuclear matter and supernova matter, which
contains leptons, at high densities (above $\rho_0$) at finite temperature
in the RMF theory have been discussed in the same references.

\section{Thomas-Fermi method for non-uniform matter}

For the density below $10^{14}g/cm^3$, where heavy nuclei can be formed
together with free nucleon gas in order to lower the free energy,
we perform the Thomas-Fermi calculation based on the work done
by Oyamatsu \cite{O93}. In this case, the system is inhomogeneous as
modeled by a mixture of electrons, muons, free nucleons,
and a single species of heavy nuclei.
The leptons can be treated as uniform non-interacting relativistic particles.
They will play no role in the minimization of the free energy for the system
with fixed proton fraction. So we mainly pay attention to baryon contribution
in this calculation.

As in Ref.\cite{O93}, we assume that each heavy spherical nucleus is
considered to be in the center of a charge-neutral cell consisting of a
vapor of neutrons, protons and leptons. The nuclei form the
body-centered-cubic (BCC) lattice to minimize the Coulomb lattice energy.

It is useful to introduce the Wigner-Seitz cell to simplify the energy of
the unit cell.
The Wigner-Seitz cell is a sphere whose volume is
the same as the unit cell in the BCC lattice.
We define the lattice constant $a$ as the cubic
root of the cell volume. Then, we have
\begin{equation}
V_{cell}=a^3=N_B / n_B,
\end{equation}
where $N_B$ and $n_{B}$ are the
baryon number per cell and the average baryon number density, respectively.
We calculate the Coulomb energy using this Wigner-Seitz approximation and
add a correction energy for the BCC lattice \cite{O93}. This correction
energy is negligible unless the nucleus size is comparable to the cell size.

We assume the nucleon distribution functions $n_i(r)$ for neutrons $(i=n)$
and protons $(i=p)$ as
\begin{equation}
n_i\left(r\right)=\left\{
\begin{array}{ll}
\left(n_i^{in}-n_i^{out}\right) \left[1-\left(\frac{r}{R_i}\right)^{t_i}
\right]^3 +n_i^{out},  & 0 \leq r \leq R_i \\
n_i^{out},  & R_i \leq r \leq R_{cell} \\
\end{array} \right. ,
\end{equation}
where $r$ represents the distance from the center of the nucleus and
$R_{cell}$ is defined by the relation,
\begin{equation}
V_{cell} \equiv \frac{4 \pi}{3} R_{cell}^3.
\end{equation}
The density parameters
$n_i^{in}$ and $n_i^{out}$ are
the densities at $r=0$ and $ r \geq R_i $, respectively. The parameters
$R_i$ and $t_i$ determine the boundary and the relative surface thickness
of the nucleus.
In order to illustrate these parameters, we show the neutron and the
proton distributions in Fig.1 for the case at
$T=10 MeV$, $Y_p=0.3$, and $\rho_B=10^{13.5}g/cm^3$.
Here we define the baryon mass density as $\rho_B=m_u n_B$ with
$m_u$ being the atomic mass unit.
For the system with fixed temperature $T$, proton fraction $Y_p$,
and baryon mass density $\rho_B$,  there are seven independent parameters
in the nine variables; $a, n_n^{in}, n_n^{out}, R_n, t_n,
n_p^{in}, n_p^{out}, R_p, t_p.$
The thermodynamically favorable state will be the one that minimizes
the free energy density with respect to these seven parameters.
In zero temperature case, there is no free proton outside nucleus,
so the number of the independent parameters is reduced to six.

In this model the free energy density contributed by baryons is given as
\begin{equation}
f\,=\,\left(\,E\,-\,T\,S\,\right)\,/\,a^3   ,
\end{equation}
where the energy per cell $E$ can be written as
\begin{equation}
E=E_{bulk}+E_s+E_C.
\end{equation}
The bulk energy $E_{bulk}$ and the entropy per cell $S$ are calculated by
\begin{eqnarray}
E_{bulk} &=&\int_{cell} \varepsilon_{RMF} \left( \, n_n\left(r\right),
\, n_p\left(r\right) \, \right) d^3r, \\
S &=&\int_{cell} s_{RMF} \left(\, n_n\left(r\right),
                            \, n_p\left(r\right) \, \right)
d^3r,
\end{eqnarray}
where $\varepsilon_{RMF}$ and $s_{RMF}$ are the energy density and the
entropy density
in the RMF theory
as functionals of the neutron density $n_n$ and the proton density $n_p$.
Note that $\varepsilon_{RMF}$ and $s_{RMF}$ at each radius
are calculated in the RMF theory for uniform nuclear matter with
the corresponding densities $n_p$ and $n_n$.
As for the surface energy term $E_s$ due to the inhomogeneity of nucleon
distribution, we take a simple form as,
\begin{equation}
E_s=\int_{cell} F_0 \mid \nabla \left( \, n_n\left(r\right)+
    n_p\left(r\right) \, \right) \mid^2 d^3r.
\end{equation}
The paramter $F_0=70 \, MeV\cdot fm^5$ is determined by doing the Thomas-Fermi
calculations of finite nuclei so as to reproduce the gross properties of
nuclear mass, charge radii and the beta stability line as described in
the appendix in Ref.\cite{O93}.
We adopt the same expression for the Coulomb energy term $E_C$ as Eq.(3.8)
in the same reference \cite{O93}.

We perform the minimization of the free energy for each density, temperature
and proton fraction in the wide range necessary for the astrophysical use.
The phase transition from non-uniform matter to uniform matter occurs
when the free energy density can not be reduced by making heavy nuclei.

\section{Equation of state of nuclear matter}

We construct the EOS of nuclear matter in the wide range of
the baryon mass density ($\rho_B=10^5 \sim 10^{15.5} g/cm^3$),
temperature ($T=0 \sim 50 MeV$), and proton fraction ($Y_p=0.01 \sim 0.5$).
We treat the uniform matter and non-uniform matter consistently using
the same RMF theory. Hence, all the resulting thermodynamical quantities
are consistent and smooth in the whole range in this set of the EOS.
We note that the EOS of nuclear matter with $Y_p=0.0$ (i.e. neutron
matter) for various densities and temperatures in the above range is
also calculated assuming the uniform distribution in all conditions.

We discuss first the calculated properties of uniform matter.
We show in Fig.2 the energy per baryon of homogeneous nuclear matter
at zero temperature as a function of the baryon density,
which is the basis of this work.
We compare the results of the RMF theory (solid curves) with
those obtained using the parameterized form based on the non-relativistic
framework in reference \cite{LS91} (dashed curves).
It is shown that the RMF theory provides the strong density dependent repulsion
over the non-relativistic results.
The calculation of free energy of uniform matter at finite temperature
has been also done in the same manner.  
Note that the free energy at zero temperature is equal to the internal energy.
Calculated properties of homogenous nuclear matter at finite temperature
have been discussed in Ref. \cite{ST94AJ,SKT95}.
The free energy per baryon of homogeneous nuclear matter is used as an input
in the Thomas-Fermi calculation.

By making the Thomas-Fermi calculations and comparing the free
energies of uniform and non-uniform configurations, we determine the
most favorable state of nuclear matter at each condition of density,
proton fraction and temperature.
In zero temperature case, the thermodynamically favorable state 
for $\rho_B \leq 10^{14} g/cm^3$
with $Y_p \geq 0.3$ is to have a BCC lattice of heavy nuclei without
outside neutron gas.
When the proton fraction $Y_p$ decreases to around $0.3$, 
the neutrons drip out from the nucleus.
In Fig.3 we show the neutron and the proton distributions along the straight
line joining the centers of the nearest nuclei in the BCC lattice 
for the case at $T=0 MeV$, $Y_p=0.4$, and 
$\rho_B=10^{12}, 10^{13}, 10^{14}, 10^{14.2} g/cm^3$.
The phase transition from  non-uniform matter to uniform matter takes
place around $\rho_B \sim 10^{14} g/cm^3$.
In this paper, we assume only the spherical nuclear shape and neglect
non-spherical
nuclear shapes, which may further reduce the free energy just before the
transition to uniform matter. However, these shapes would not give
siginificant
change to the EOS of matter but only make the transition smoother.
The non-spherical shapes can reduce slightly the surface
and Coulomb energy compared with the spherical shape, when the size of the
nucleus
is comparable to the cell size as in the case of  
$\rho_B=10^{14} g/cm^3$ in Fig.3. However,
in these cases, the surface and Coulomb energies are much smaller than the
bulk energy so that the energy gain due to the shape change is relatively
small.
We shall calculate the nuclear mass number, $A=Z+N$, in the higher
desity part of the inhomogeneous nuclear matter by integrating
the proton number, $Z$, and the neutron number, $N$, up to the radius
$R_{max}$, which is the maximum between $R_p$ and $R_n$.
Here $R_{max}$ is considered as the boundary of the nucleus.
In Figures 4a and 4b, we display the nuclear mass number $A$ and the proton
number $Z$
as a function of the baryon mass density $\rho_B$ with various proton
fraction $Y_p$.
Both $A$ and $Z$ increase sharply just before the
phase transition occurs, reflecting that the nucleus becomes larger and
the system turns into the uniform matter phase.

In finite temperature case, the second term in Eq.(6) has
appreciable contribution to the free energy density as the temperature
increases.
We show the phase diagram of non-uniform
matter and uniform matter at $T=0, 1, 5$ and $10 MeV$ in Fig.5.
For extremely low density and finite temperature range,
the thermodynamically favorable state is uniform nucleon gas.
We can approximate the uniform nucleon gas
as a classical non-interacting gas below $\rho_B < 10^{10} g/cm^3$.
The phase transition from uniform nucleon gas to non-uniform matter
takes place at the density where the system can lower the free energy
by making heavy nuclei. It is seen in Fig.5  that the lower phase
transition density (from uniform nucleon gas to non-uniform matter)
depends on the temperature very strongly, while the higher phase
transition density (from non-uniform matter to uniform matter)
is nearly independent of the temperature, which is found almost constant
at $\rho_B \sim 10^{14} g/cm^3$.
As the temperature increases, the  area of the non-uniform matter phase
becomes smaller. The non-uniform matter phase disappears
when the temperature is higher than $\sim 15 MeV$.
We show in Fig.6 the neutron and the proton distributions for the case at
 $\rho_B=10^{13.5} g/cm^3$, $Y_p=0.4$, and  $T=0, 5, 10, 15 MeV$.
From this figure, we can find how the properties of the Wigner-Seitz
cell change as the temperature increases.
The neutron and the proton densities outside the nucleus ( $n_n^{out}$
and $n_p^{out}$ ) become  large numbers at high temperature,
while $n_p^{out}=0$ at $T=0 MeV$ for any $\rho_B$ and $Y_p$.
There is a tendency that the nucleon number ratio between inside and outside
the heavy nucleus decreases as the temperature increases.
Furthermore, the heavy nuclei are dissociated when the temperature
is high enough.

Fig.7 shows the free energy per baryon $f/n_B$ as a function of
the baryon mass density $\rho_B$ with various $Y_p$ at $T=0, 1, 5$ and $10
MeV$.
In the uniform nucleon gas phase at low density and finite temperature
regime,
the $f/n_B$ are not significantly altered by variations in $Y_p$,
because the interaction between nucleons is very weak at low density.
When the system changes into the non-uniform matter phase,
the formation of heavy nuclei can reduce  $f/n_B$,
and brings appreciable $Y_p$ dependence of $f/n_B$.
All curves display a rapid rise in the high density range, which is
due to the strong density dependent repulsion among nucleons in the RMF theory.
We show the results of the entropy per baryon as a function of
$\rho_B$ with various $Y_p$ at $T=1, 5$ and $10 MeV$ in Fig.8.
The effect of the non-uniform matter phase on the entropy clearly
appear in this figure.
The drop of the entropy in the non-uniform matter region is due to
the formation of heavy nuclei. The $Y_p$ dependence of the entropy in
the non-uniform matter region is opposite to the one in the
uniform matter regions when the temperature is low. This is because
more heavy nuclei are formed in non-uniform matter phase at higher $Y_p$,
then they can bring more rapid drop in the entropy per baryon curves
comparing with the case at lower $Y_p$.
On the other hand, the free nucleon gas outside is a dominant composition
of matter
at high temperature, and the effect of the formation of heavy nuclei
is weakened as the temperature increases.

The pressure can be calculated from $f/n_B$ by the following
thermodynamical equilibrium condition,
\begin{equation}
P= n_B^2 \, \frac {\partial \left( f/n_B \right) }
 { \partial n_B} \mid_{T,Y_p}.
\end{equation}
We show in Fig.9 the pressure as a function of $\rho_B$
with various $Y_p$ at $T=0, 1, 5$ and $10 MeV$.
In the uniform nucleon gas phase at low density and at finite temperature,
the pressure is mainly provided by the thermal pressure, which is given by
$P=n_B\, T$. In this case, the pressure is independent of $Y_p$.
When the system goes into the non-uniform matter phase,
the reduction of $f/n_B$
due to the formation of heavy nuclei brings a sharp drop in pressure.
For high $Y_p$, the pressure drops to the negative area.
The drop of the pressure in the non-uniform matter phase has  a strong
 $Y_p$ dependence.
In high density region, all curves come back to the positive area,
and have a rapid rise as the densiy increases.
The behavior of the pressure at high density is determined
predominantly by the contribution of the vector meson, $\omega$,
and isovector-vector meson, $\rho$, in the RMF theory \cite{SKT95}.
Because of the non-linear term of the $\omega$ meson in Eq.(1),
the repulsive $\omega$ meson field is suppressed at high density.
So it does not increase linearly in density as those of another
RMF theory without this non-linear $\omega$ meson term.
As a result, the pressure is reduced largely at high density
as compared with the case in the other RMF theory.
We stress that this behavior is accord with that in the relativistic
Br\"uckner-Hartree-Fock (RBHF) theory, on which the RMF theory
is based.

\section{Neutron star matter and neutron star profiles}

We construct the EOS of nuclear matter with various proton fractions
so that it can be used in the simulation of supernova explosion
where the $\beta$-equilibrium may not be achieved within its timescale.
In the study of cold neutron stars, the $\beta$-equilibrium condition
can be assumed as for static properties.
In order to calculate the properties of neutron star matter at zero
temperature,
we have to incorporate the contribution of leptons in the EOS.
The proton fraction of neutron star matter, which is determined from the
$\beta$-equilibrium condition, is shown by the dot-dashed curve in Fig.5.
Large proton fractions at high densities beyond $10^{14} g/cm^3$
is remarkable in the RMF theory \cite{ST94,ST94AJ,SKT95}.
We display in Fig.10 the pressure of neutron star matter as a function
of the mass density. Here the mass density is equal to the total 
energy density divided by $c^2$.
Since we have taken account of the non-uniform matter phase,
the present EOS named NURMF (non-uniform relativistic mean field)
is largely different from the EOS named URMF (uniform relativistic mean
field), which is obtained by assuming uniform matter in the same RMF theory
for low density range \cite{SKT95}.
The EOS of BJ obtained by Bethe and Johnson \cite{BJ74}
is also shown for comparison in Fig.10 by the short-dashed curve.

Applying the EOS of neutron star matter,
we calculate the neutron star profile by solving
the Oppenheimer-Volkoff equation. The gravitational masses of neutron
star as a function of the central mass density are displayed in Fig.11.
The neutron star mass is not signficantly altered by using the EOS
of NURMF comparing with the result using the EOS of URMF.
This is because the neutron star mass is determinded predominantly by
the behavior of the EOS at high density. On the other hand,
the properties of the EOS at low density are important in the description
of the neutron star profile in the surface region,
where the matter is inhomogeneous.
In order to illustrate the neutron star profile, we show the baryon mass 
density distribution of the neutron star having a gravitational mass of
$1.4 M_{\odot}$ in Fig.12.
We display the boundaries of the uniform matter and the non-uniform matter
with and without free dripping neutrons by short-dashed and
dot-dashed lines.
It is difficult to arrive at the low density region in the calculation
of the neutron star profile by using the EOS of URMF, and it is also not
appropriate to describe the neutron star matter by uniform RMF approach
as $\rho_B \leq \rho_0/3$ where heavy nuclei are formed.
In Fig.13, we expand the surface region of the Fig.12 in order to see
the difference of the neutron star profiles between NURMF and URMF
at the large radius region.
It is interesting to mention that the neutron star radius increases by having
the matter composed of nuclei in the most outer part of the neutron star.
For the neutron star having a gravitational mass of $1.4 M_{\odot}$,
the central baryon
density is around $1.8 \rho_0$. In this case, the proton fraction in the
central core of the neutron star is around 0.19 as can be seen by
the $\beta$-equilibrium curve in Fig.5.

\section{Conclusions and discussions}

This relativistic EOS of nuclear matter is designed for the use
of supernova simulations and the neutron star calculations
in the wide density and temperature range with various proton fractions.
We have adopted the RMF
theory with the non-linear $\sigma$ and $\omega$ terms, which was
demonstrated to be successful in describing the properties of both
stable and unstable nuclei as well as reproducing the similar
self-energies in the RBHF theory. The relativity plays an essential
role in describing the nuclear saturation and the nuclear
structure \cite{BM90}, it also brings some distinctive properties
in the EOS comparing with the case in the non-relativistic framework.
Therefore, it is very interesting and important
to study the astrophysical phenomena
such as supernova explosion, neutron star cooling and neutron star
merger using the relativistic EOS through extensive comparison with the
previous
studies with the non-relativistic EOS.

We have performed the Thomas-Fermi calculation to describe the non-uniform
matter which appears below $\rho_B \sim \rho_0/3$. The same RMF theory
as the one used at high density has been adopted as the input of
the Thomas-Fermi numerical code. Hence, all the thermodynamical quantities
such as free energy, entropy and pressure are found smooth in the
resulting EOS.
The non-uniform matter phase appears in a wide $\rho_B-Y_p$ region
in the phase diagram at low temperature,
while it completely disappears as temperature increases beyond $15 MeV$.
The formation of heavy nuclei in non-uniform matter  provides
remarkable behavior in the EOS.

It is also important to pay attention to the experimental data of unstable
nuclei, which are being obtained using radioactive nuclear beams,
so that we can get more information about the properties of
asymmetric nuclear matter from experiments \cite{O98}.
It is definitely important and meaningful to study unstable nuclei and
astrophysical phenomena simultaneously in the same framework.

\section*{Acknowledgment}

We would like to thank S.Yamada and H.Suzuki for fruitful discussions.
H.S. acknowledges the COE program for enabling her to stay at RCNP-Osaka,
where this work was carried out.  A large portion of the present calculation
has been done using the supercomputers at RIKEN and RCNP.


\section*{Figure captions}

\begin{description}
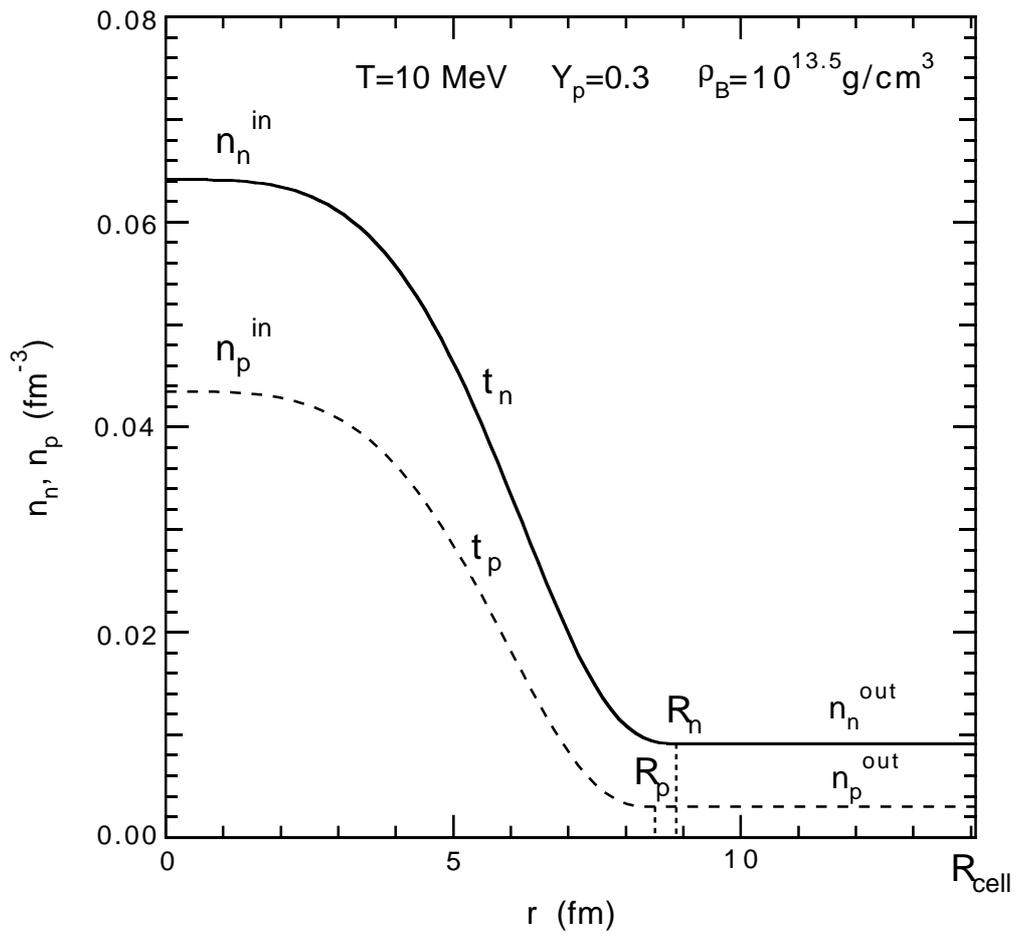
\item[Fig.1.] The neutron distribution and the proton distribution in the
Wigner-Seitz cell at $T=10 MeV$, $Y_p=0.3$, and $\rho_B=10^{13.5}g/cm^3$
are shown by solid curve and dashed curve, respectively.
The parameters $n_i^{in}, n_i^{out}, R_i$ and $t_i$
of the density distributions for protons and neutrons
are defined in Eq.(4), while $R_{cell}$ represents the radius of
spherical Wigner-Seitz cell.

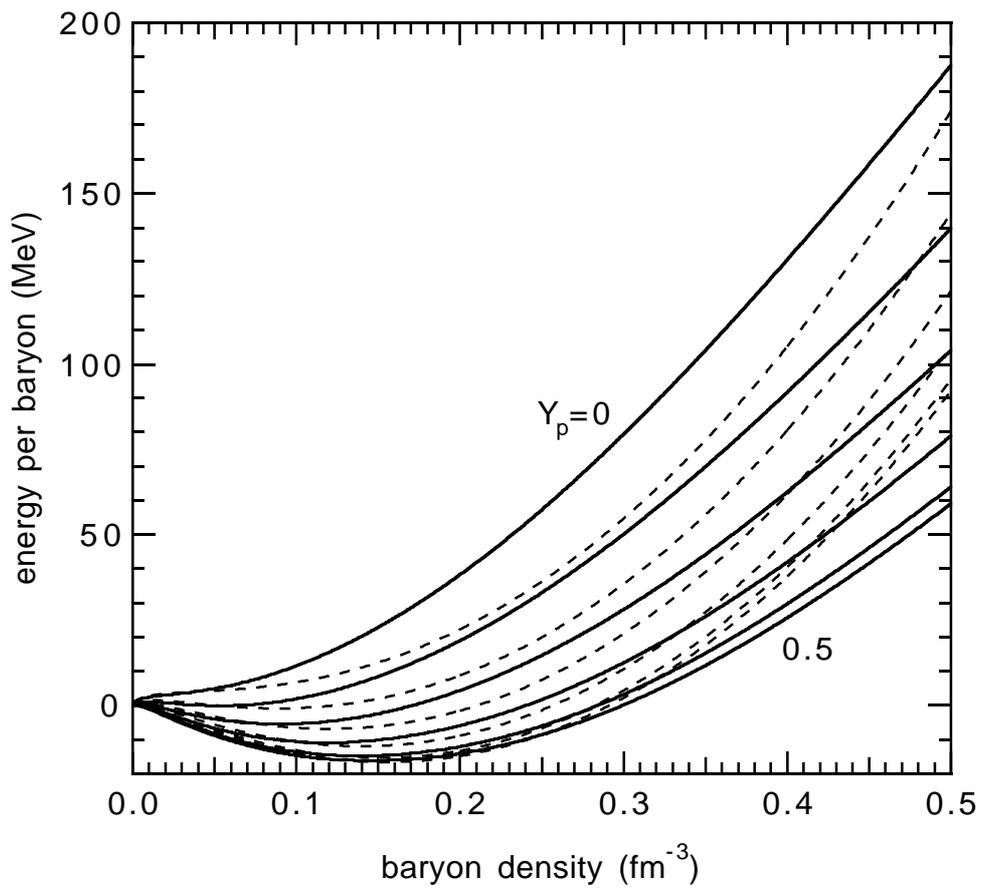
\item[Fig.2.] The energy per baryon as a function of the baryon density
for homogeneous nuclear matter at zero temperature with the proton fraction
$Y_p=0, 0.1, 0.2, 0.3, 0.4,$ and $0.5$. The solid curves
represent the results of the RMF theory, while the non-relativistic
results taken from \cite{LS91} are shown by dashed curves for comparison.

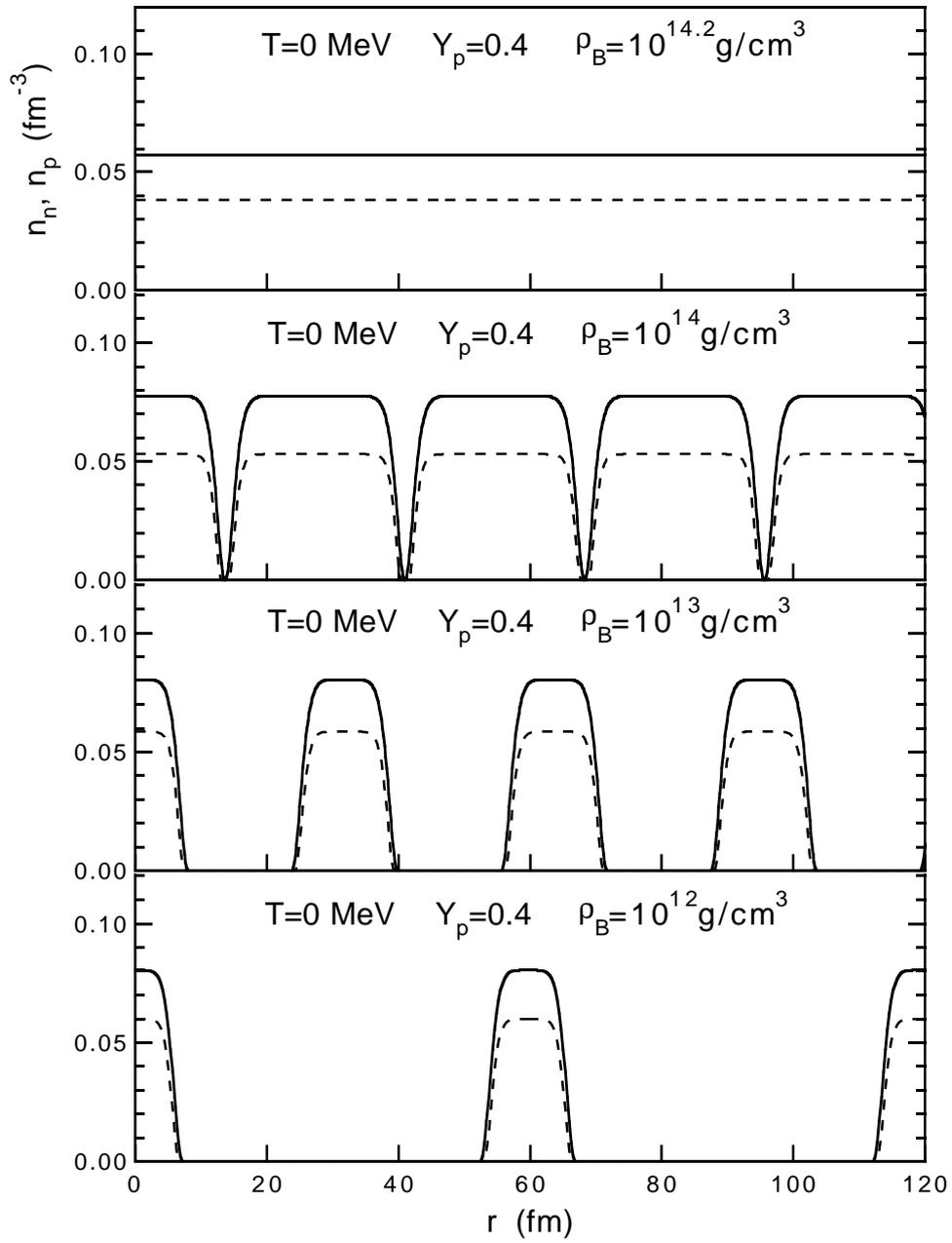
\item[Fig.3.] The neutron distribution (solid curves) and the proton
distribution (dashed curves)
along the straight lines joining the centers of the 
nearest nuclei in the BCC lattice for the case 
at $T=0 MeV$, $Y_p=0.4$, 
and $\rho_B=10^{12}, 10^{13}, 10^{14}, 10^{14.2} g/cm^3$.

\item[Fig.4a.] The nuclear mass number $A$ as a function of
the baryon mass density $\rho_B$ with the proton fraction
$Y_p=0.1, 0.2, 0.3, 0.4,$ and $0.5$ at zero temperature.

\item[Fig.4b.] The nuclear proton number $Z$ as a function of
the baryon mass density $\rho_B$ with the proton fraction
$Y_p=0.1, 0.2, 0.3, 0.4,$ and $0.5$ at zero temperature.

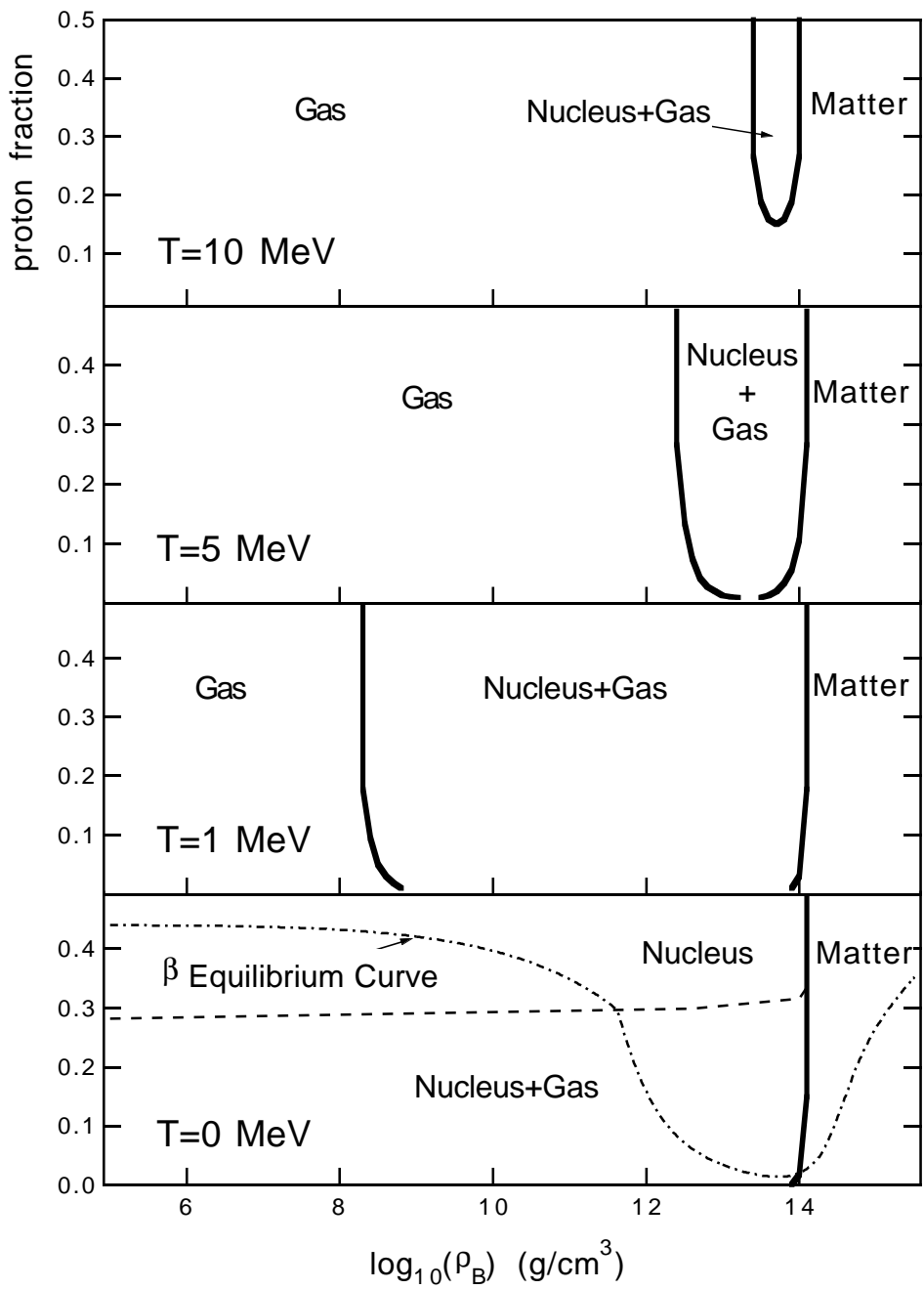
\item[Fig.5.] The phase diagrams of the nuclear matter at
$T=0, 1, 5,$ and $10 MeV$ in $\rho_B - Y_p$ plane.
The regions denoted by Matter correspond
to the uniform matter described by RMF theory.
The regions denoted by Nucleus+Gas and Nucleus correspond to the
non-uniform matter with and without free nucleon gas, the boundary
between them is shown by dashed curve at $T=0 MeV$.
The regions denoted by Gas correspond to the uniform matter close to the
classical ideal gas approximation. The $\beta$-equilibrium curve is also
shown at $T=0 MeV$ by dot-dashed curve.

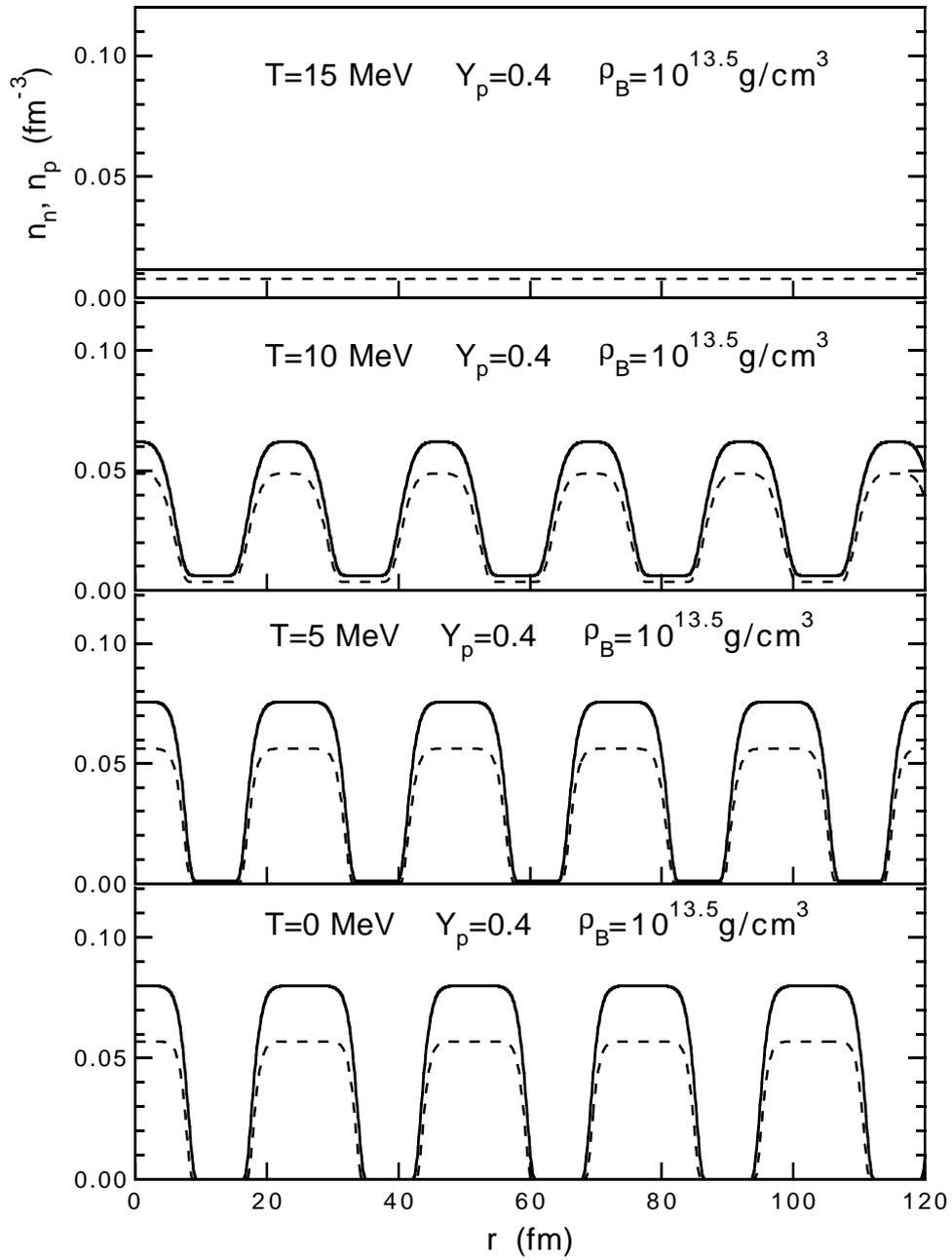
\item[Fig.6.] The neutron distribution (solid curves) and the proton
distribution (dashed curves)
along the straight lines joining the centers of the 
nearest nuclei in the BCC lattice for the case 
at $\rho_B=10^{13.5} g/cm^3$, $Y_p=0.4$, and $T=0, 5, 10, 15 MeV$.

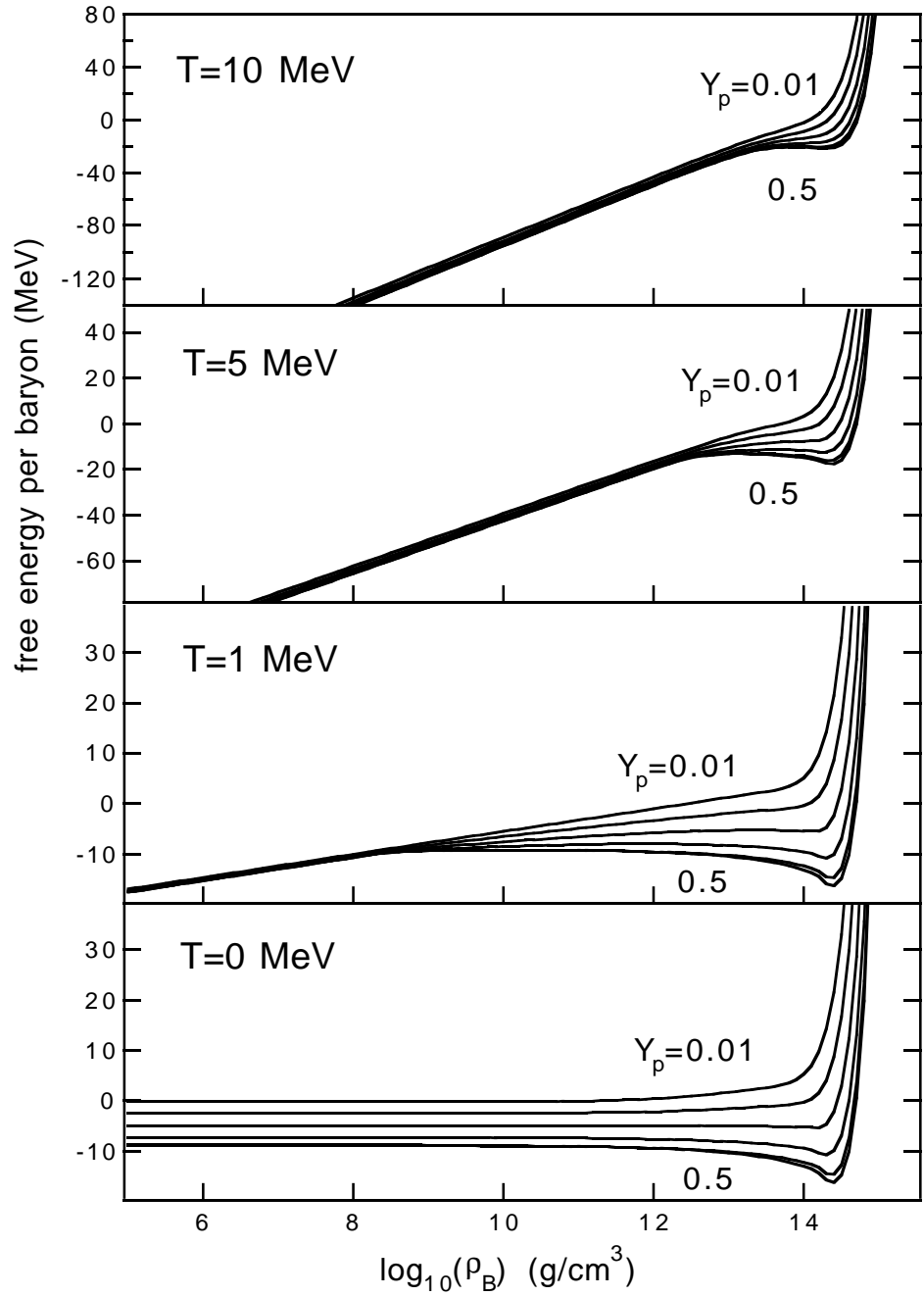
\item[Fig.7.] The free energy per baryon as a function of the baryon
mass density
$\rho_B$ with the proton fraction $Y_p=0.01, 0.1, 0.2, 0.3, 0.4,$ and $0.5$
at $T=0, 1, 5,$ and $10 MeV$.

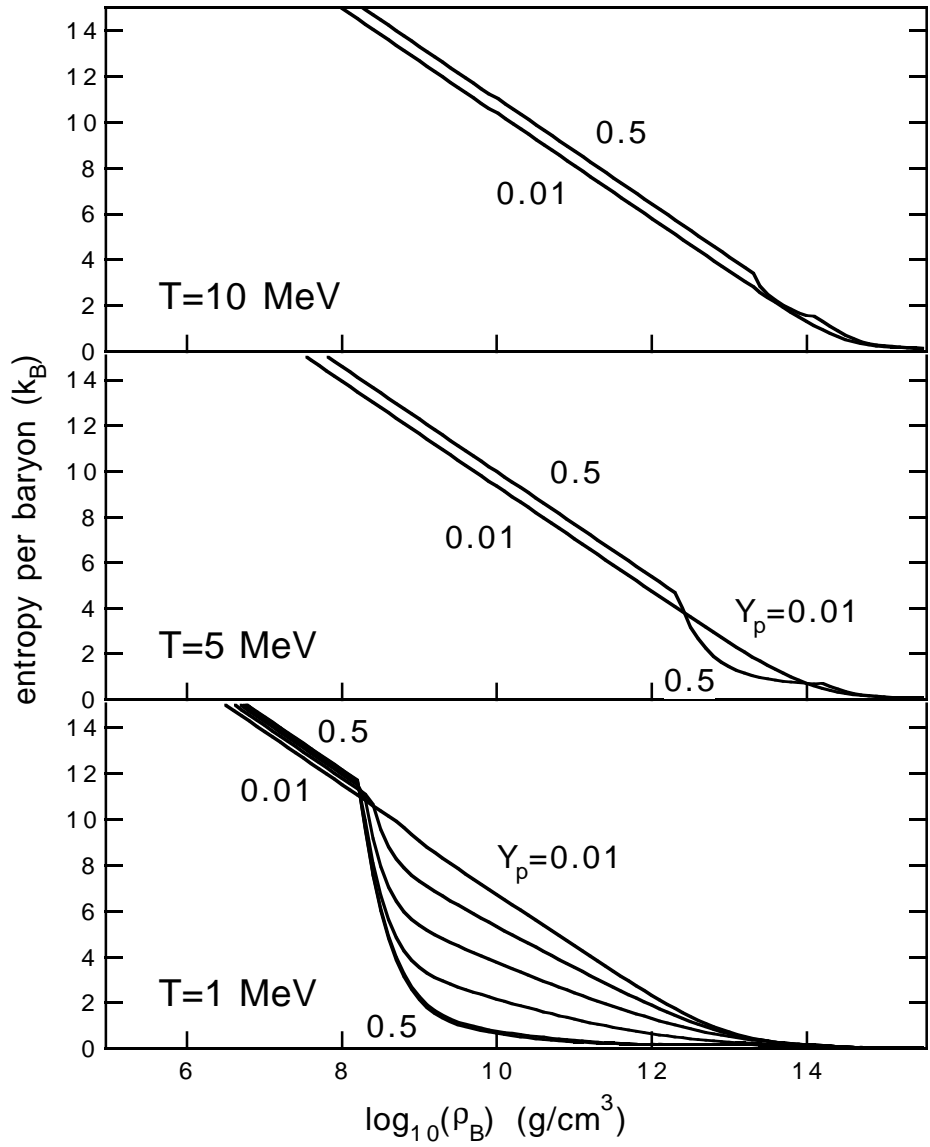
\item[Fig.8.] The entropy per baryon as a function of the baryon mass density
$\rho_B$ with the proton fraction $Y_p=0.01, 0.1, 0.2, 0.3, 0.4,$ and $0.5$
at $T=1 MeV$, and with $Y_p=0.01$ and $0.5$ at $T=5$ and $10 MeV$.

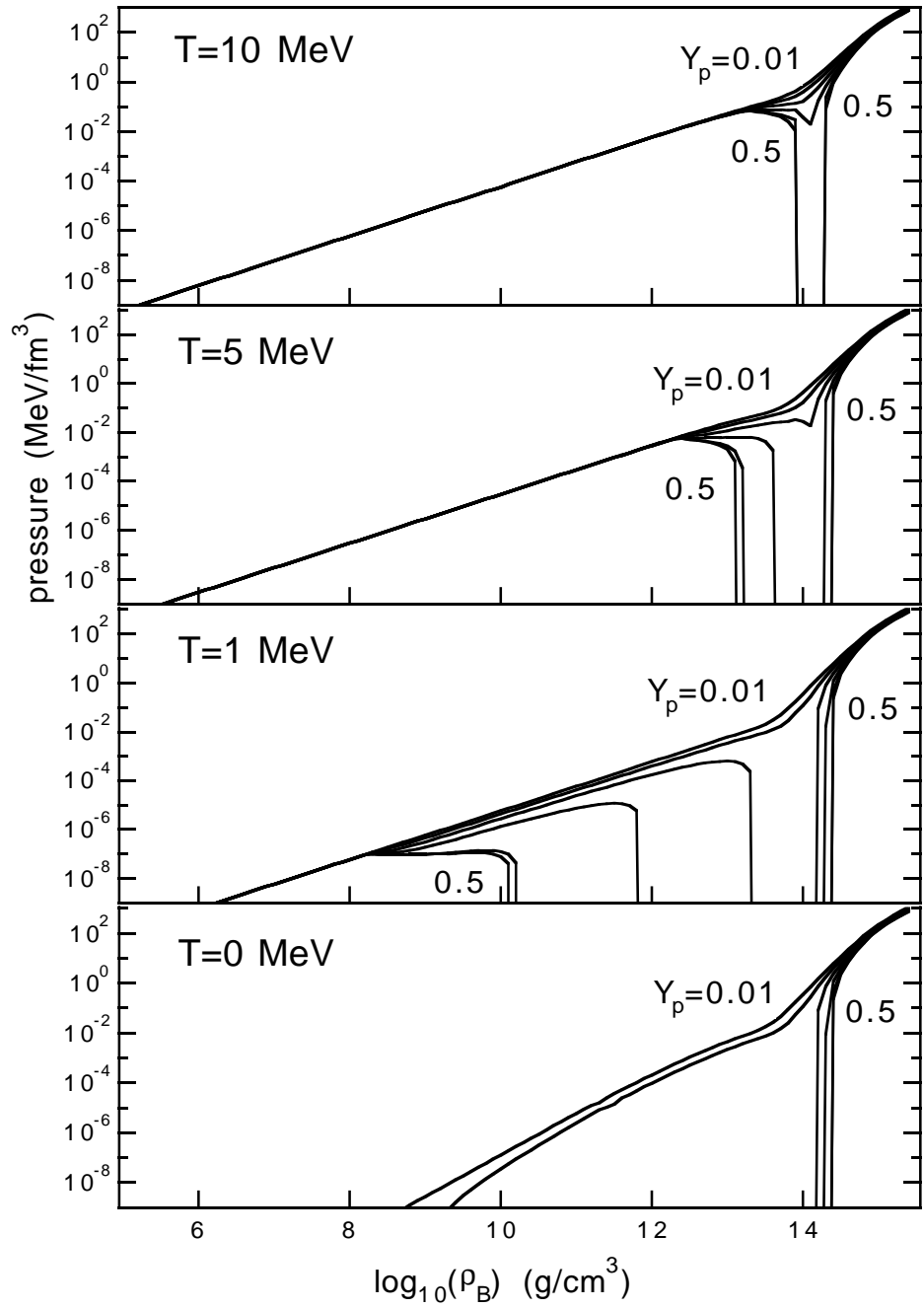
\item[Fig.9.] The pressure as a function of the baryon mass density
$\rho_B$ with the proton fraction $Y_p=0.01, 0.1, 0.2, 0.3, 0.4,$ and $0.5$
at $T=0, 1, 5,$ and $10 MeV$.

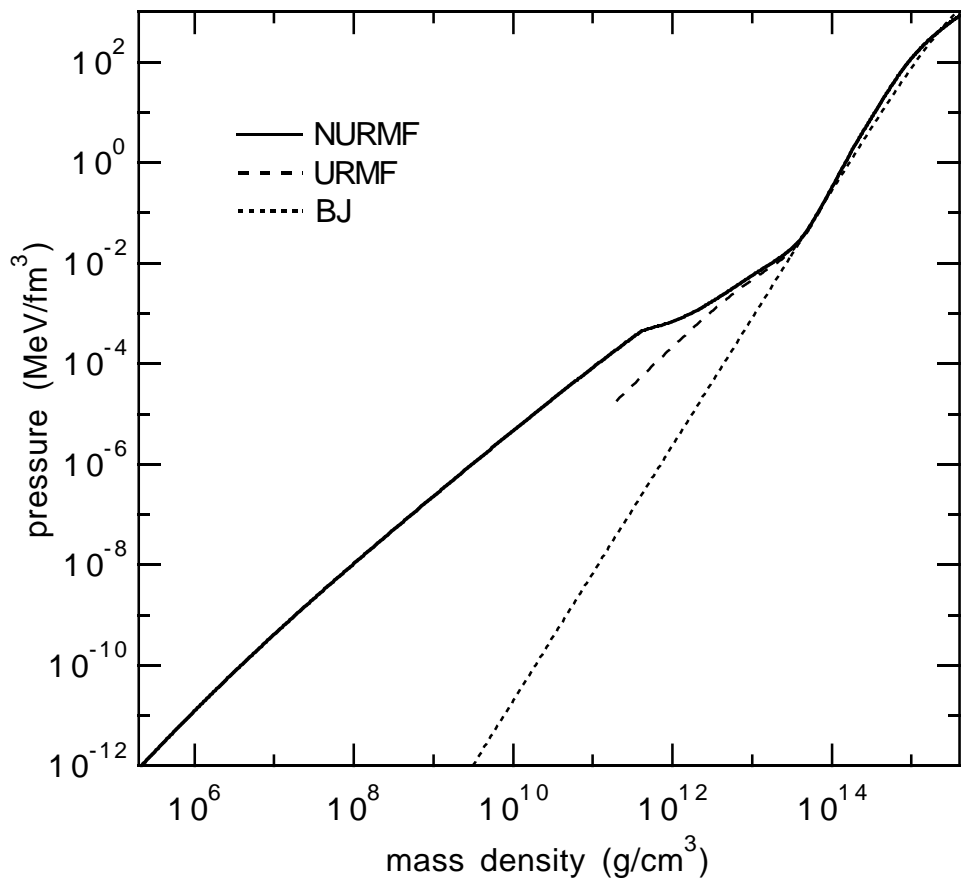
\item[Fig.10.] The pressure of neutron star matter at zero temperature
is shown as a function of the mass density.
The solid curve is the EOS of NURMF including the non-uniform matter phase
at low density range, while the dashed curve is the EOS of URMF
done by just assuming uniform matter phase in the same RMF theory.
The EOS of BJ is also shown by short-dashed curve for comparison.

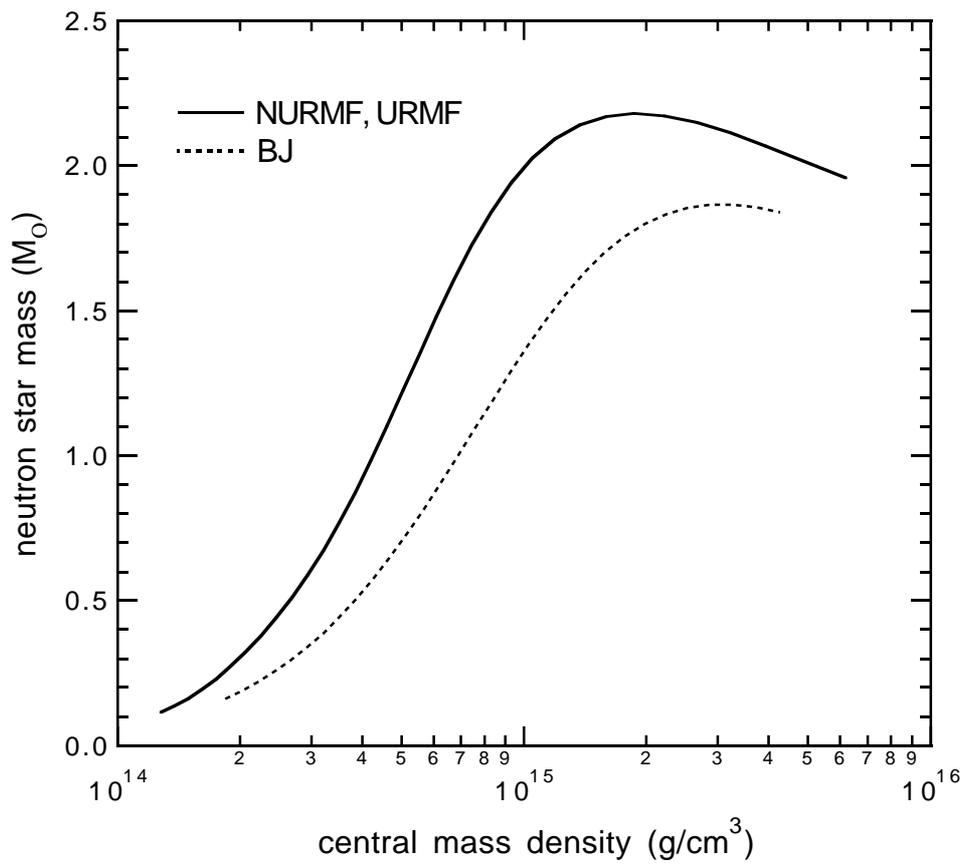
\item[Fig.11.] The gravitational masses of neutron star with the EOS of
NURMF and URMF are shown by the solid curve as a function
of the central mass density. There is no obvious difference between
the results using the EOS of NURMF and URMF. The neutron
star masses calculated with the EOS of BJ are shown by the
short-dashed curve for comparison.

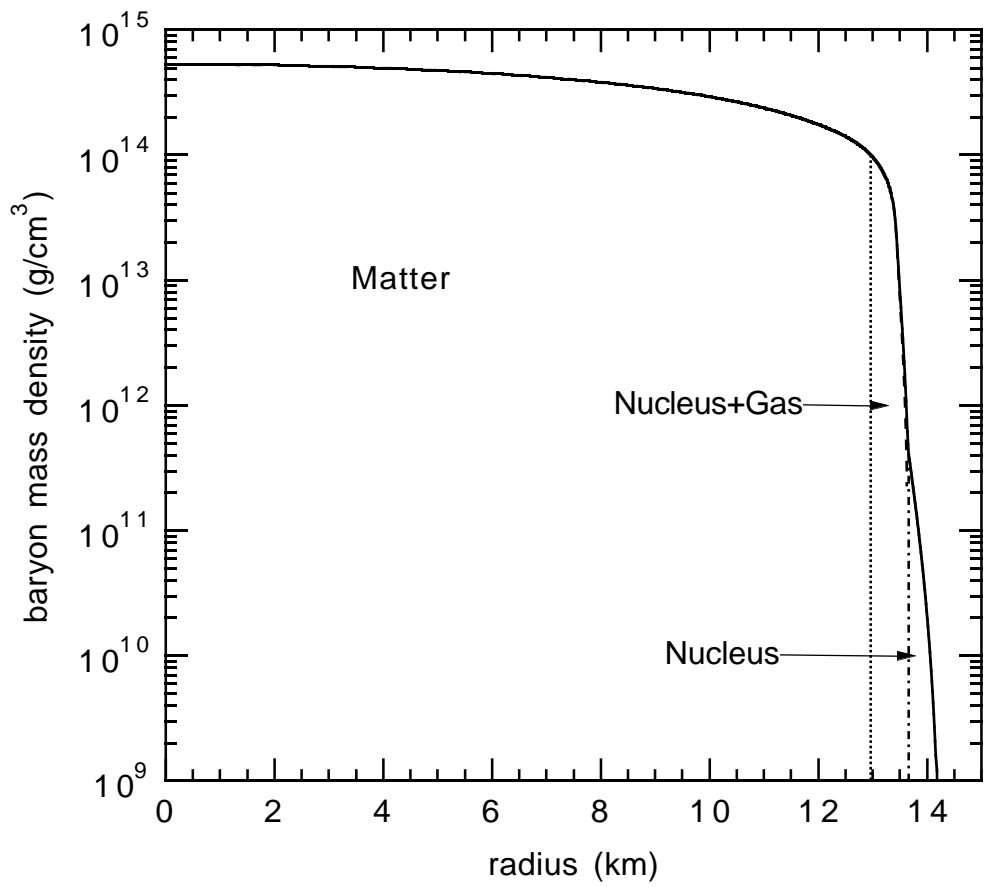
\item[Fig.12.] The baryon mass density profiles of the neutron star having
a gravitational mass of $1.4 M_{\odot}$ using the EOS of NURMF and
URMF are shown by solid curve and dashed curve, respectively.
We display the boundaries of the uniform matter and the non-uniform matter
with and without free dripping neutrons by short-dashed and
dot-dashed lines. There is no obvious difference between
the results using the EOS of NURMF and URMF in the central
core of the neutron star. The difference exists in the surface region
as can be seen in Fig. 13.

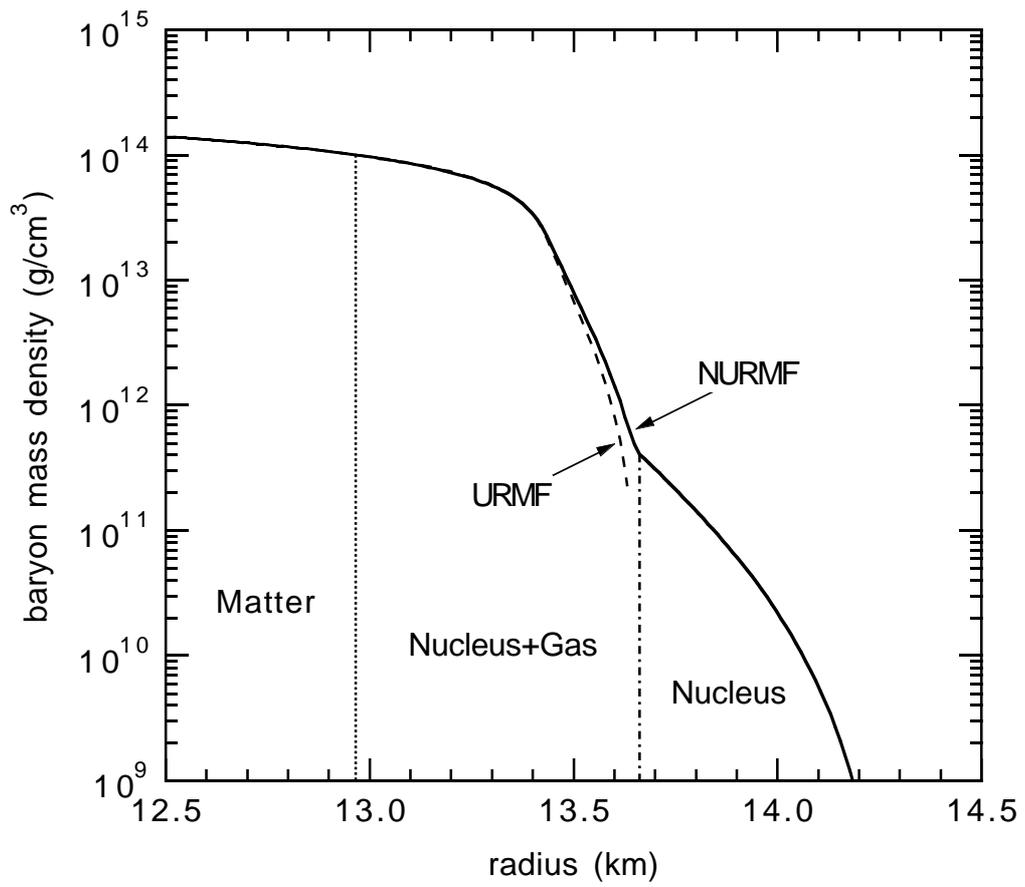
\item[Fig.13] The expansion of Fig.12 in the surface region.
The difference between the neutron star profiles with the EOS of NURMF
and URMF is seen at the large radius region.

\end{description}

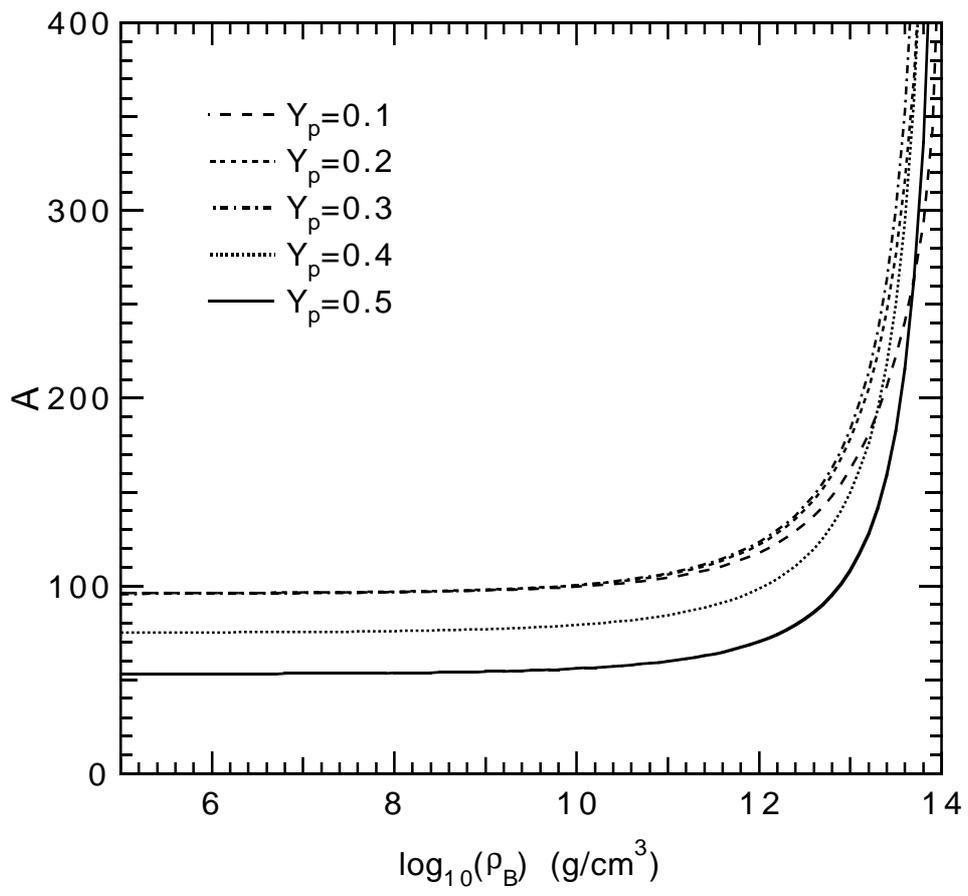
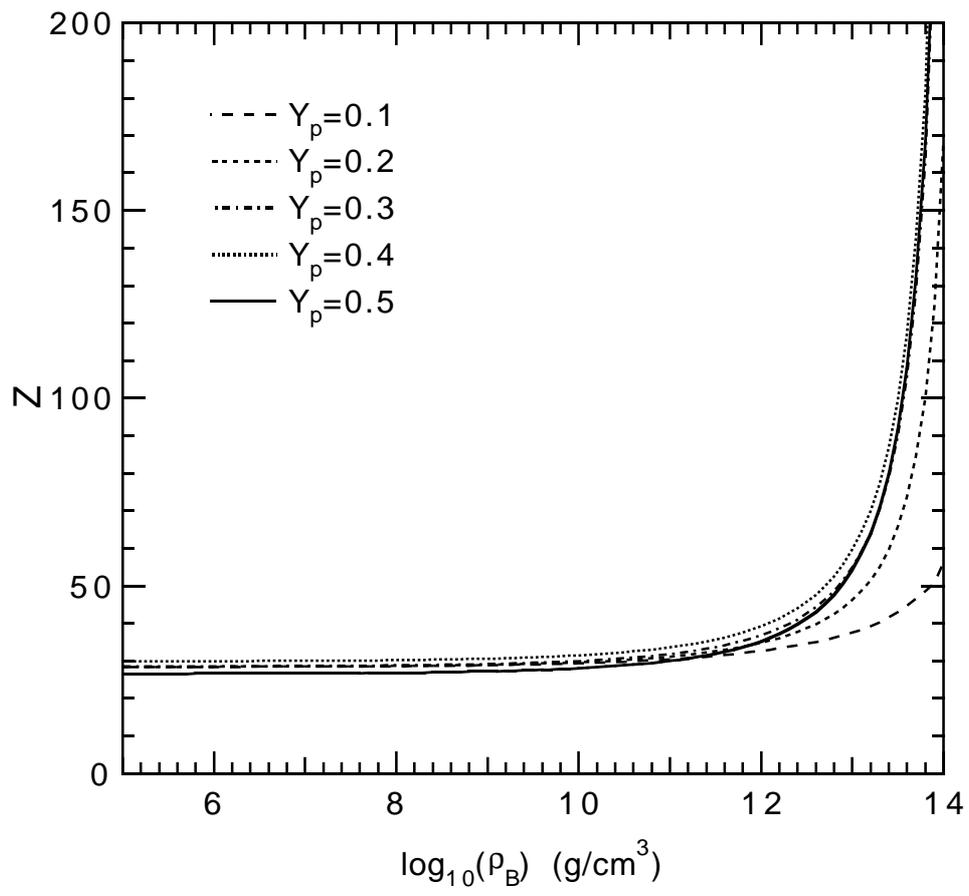
\end{document}